\documentclass{emulateapj}
\usepackage{apjfonts}
\usepackage{rotating}

\newcommand{\pivec}{\mbox{\boldmath $\pi$}}

\lefthead{CHOI ET AL.} 
\righthead{SOURCE-TRANSIT LENSING EVENTS}

\begin{document}
\title{Characterizing Lenses and Lensed Stars of
High-Magnification Single-lens Gravitational Microlensing Events
With Lenses Passing Over Source Stars}

\author{
J.-Y. Choi$^{1}$,
I.-G. Shin$^{1}$,
S.-Y. Park$^{1}$,
C. Han$^{1,68,74}$,
A. Gould$^{2,68}$,
T. Sumi$^{3,69}$,        
A. Udalski$^{4,70}$,    
J.-P. Beaulieu$^{5,71}$, 
R. Street$^{6,72}$,       
M. Dominik$^{7,73}$\\
and\\
W. Allen$^{8}$,
L.A. Almeida$^{67}$,
M. Bos$^{9}$,
G.W. Christie$^{10}$,   
D.L. Depoy$^{11}$,      
S. Dong$^{12}$, 
J. Drummond$^{13}$,
A. Gal-Yam$^{14}$,
B.S. Gaudi$^{2}$,\\      
C.B. Henderson$^{2}$,
L.-W. Hung$^{15}$,
F. Jablonski$^{67}$,
J. Janczak$^{16}$,
C.-U. Lee$^{17}$,      
F. Mallia$^{18}$,
A. Maury$^{18}$,
J. McCormick$^{19}$,\\    
D. McGregor$^{2}$,
L.A.G. Monard$^{20}$,
D. Moorhouse$^{21}$,    
J. A. Mu\~{n}oz$^{22}$,
T. Natusch$^{10}$,
C. Nelson$^{23}$,
B.-G. Park$^{17}$, 
R.W.\ Pogge$^{2}$,\\      
T.-G. "TG" Tan$^{24}$,
G. Thornley$^{21}$,
J.C. Yee$^{2}$\\
(The $\mu$FUN Collaboration),\\
F. Abe$^{25}$,        
E. Barnard$^{26}$,   
J. Baudry$^{26}$,
D.P. Bennett$^{27}$,  
I.A. Bond$^{28}$,
C.S. Botzler$^{26}$,   
M. Freeman$^{26}$,
A. Fukui$^{29}$,\\   
K. Furusawa$^{25}$,   
F. Hayashi$^{25}$,    
J.B. Hearnshaw$^{30}$,
S. Hosaka$^{25}$,     
Y. Itow$^{25}$,        
K. Kamiya$^{25}$,      
P.M. Kilmartin$^{31}$, 
S. Kobara$^{25}$,\\
A. Korpela$^{32}$,     
W. Lin$^{28}$,         
C.H. Ling$^{28}$,     
S. Makita$^{25}$,
K. Masuda$^{25}$,      
Y. Matsubara$^{25}$,    
N. Miyake$^{25}$,
Y. Muraki$^{33}$,     
M. Nagaya$^{25}$,\\
K. Nishimoto$^{25}$,   
K. Ohnishi$^{34}$,    
T. Okumura$^{25}$,
K. Omori$^{25}$,
Y.C. Perrott$^{26}$,   
N. Rattenbury$^{26}$,   
To. Saito$^{35}$,    
L. Skuljan$^{28}$,\\     
D.J. Sullivan$^{32}$,  
D. Suzuki$^{3}$,
K. Suzuki$^{25}$,
W.L. Sweatman$^{28}$, 
S. Takino$^{25}$,
P.J. Tristram$^{31}$,  
K. Wada$^{3}$,        
P.C.M. Yock$^{26}$\\   
(The MOA Collaboration),\\
M.K. Szyma\'nski$^{4}$,  
M. Kubiak$^{4}$,            
G. Pietrzy\'nski$^{4,36}$, 
I. Soszy\'nski$^{4}$,      
R. Poleski$^{4}$,
K.\ Ulaczyk$^{4}$,
{\L}. Wyrzykowski$^{4,37}$,\\
S. Koz{\l}owski$^{4}$,
P. Pietrukowicz$^{4}$\\
(The OGLE Collaboration)\\
M.D. Albrow$^{30}$,
E. Bachelet$^{38}$,
V. Batista$^{2}$,    
C.S. Bennett$^{39}$,
R. Bowens-Rubin$^{40}$,
S. Brillant$^{41}$,     
A. Cassan$^{5}$,       
A. Cole$^{42}$,\\         
E. Corrales$^{5}$,      
Ch. Coutures$^{5}$,
S. Dieters$^{5,38}$,    
D. Dominis Prester$^{43}$,
J. Donatowicz$^{44}$,
P. Fouqu\'e$^{38}$,     
J. Greenhill$^{42}$,\\     
S. R. Kane$^{45}$,
J. Menzies$^{46}$,
K. C. Sahu$^{47}$,
J. Wambsganss$^{48}$,
A. Williams$^{49}$,
M. Zub$^{48}$\\
(The PLANET Collaboration)\\
A. Allan$^{50}$,
D.M. Bramich$^{51}$,
P. Browne$^{7}$, 
N. Clay$^{52}$,
S. Fraser$^{52}$,
K. Horne$^{7}$,          
N. Kains$^{51}$
C. Mottram$^{52}$,
C. Snodgrass$^{53,41}$,\\  
I. Steele$^{52}$,  
Y. Tsapras$^{6}$\\
(The RoboNet Collaboration)\\
and\\
K.A. Alsubai$^{54}$,
V. Bozza$^{55}$,                 
M.J. Burgdorf$^{56}$, 
S. Calchi Novati$^{55}$, 
P. Dodds$^{7}$,
S. Dreizler$^{57}$, 
F. Finet$^{58}$,
T. Gerner$^{48}$,\\ 
M. Glitrup$^{59}$, 
F. Grundahl$^{59}$, 
S. Hardis$^{60}$,
K. Harps{\o}e$^{60,61}$, 
T.C. Hinse$^{17,60}$,
M. Hundertmark$^{7,57}$,
U.G. J{\o}rgensen$^{60}$,\\ 
E. Kerins$^{62}$,
C. Liebig$^{48}$, 
G. Maier$^{48}$, 
L. Mancini$^{55,63}$, 
M. Mathiasen$^{60}$,
M.T. Penny$^{62}$,
S. Proft$^{48}$,
S. Rahvar$^{64,65}$, 
D. Ricci$^{58}$,\\ 
G. Scarpetta$^{55}$, 
S. Sch\"{a}fer$^{57}$, 
F. Sch\"{o}nebeck$^{48}$,
J. Skottfelt$^{60}$,
J. Surdej$^{58}$, 
J. Southworth$^{66}$, 
F. Zimmer$^{48}$\\
(The MiNDSTEp Consortium)\\
}

\bigskip\bigskip
\affil{$^{1}$Department of Physics, Institute for Astrophysics, Chungbuk National University, Cheongju 371-763, Korea}
\affil{$^{2}$Department of Astronomy, Ohio State University, 140 W. 18th Ave., Columbus, OH 43210, USA}
\affil{$^{3}$Department of Earth and Space Science, Osaka University, Osaka 560-0043, Japan}
\affil{$^{4}$Warsaw University Observatory, Al. Ujazdowskie 4, 00-478 Warszawa, Poland}
\affil{$^{5}$Institut d'Astrophysique de Paris, UMR7095 CNRS--Universit{\'e} Pierre \& Marie Curie, 98 bis boulevard Arago, 75014 Paris, France} 
\affil{$^{6}$Las Cumbres Observatory Global Telescope Network, 6740B Cortona Dr, Suite 102, Goleta, CA 93117, USA} 
\affil{$^{7}$School of Physics \& Astronomy, SUPA, University of St. Andrews, North Haugh, St. Andrews, KY16 9SS, UK}
\affil{$^{8}$Vintage Lane Observatory, Blenheim, New Zealand}
\affil{$^{9}$Molehill Astronomical Observatory, North Shore, New Zealand}
\affil{$^{10}$Auckland Observatory, P.O. Box 24-180, Auckland, New Zealand} 
\affil{$^{11}$Department of Physics, Texas A\&M University, College Station, TX, USA} 
\affil{$^{12}$Institute for Advanced Study, Einstein Drive, Princeton, NJ 08540, USA}
\affil{$^{13}$Possum Observatory, Patutahi, New Zealand}
\affil{$^{14}$Benoziyo Center for Astrophysics, the Weizmann Institute, Israel} 
\affil{$^{15}$Department of Physics \& Astronomy, University of California Los Angeles, Los Angeles, CA 90095, USA}
\affil{$^{16}$Department of Physics, Ohio State University, 191 W. Woodruff, Columbus, OH 43210, USA}
\affil{$^{17}$Korea Astronomy and Space Science Institute, Daejeon 305-348, Korea} 
\affil{$^{18}$Campo Catino Austral Observatory, San Pedro de Atacama, Chile}
\affil{$^{19}$Farm Cove Observatory, Pakuranga, Auckland} 
\affil{$^{20}$Bronberg Observatory, Pretoria, South Africa} 
\affil{$^{21}$Kumeu Observatory, Kumeu, New Zealand} 
\affil{$^{22}$Departamento de Astronom\'{\i}a y Astrof\'{\i}sica, Universidad de Valencia, E-46100 Burjassot, Valencia, Spain}
\affil{$^{23}$College of Optical Sciences, University of Arizona, 1630 E. University Blvd, Tucson Arizona, 85721}
\affil{$^{24}$Perth Exoplanet Survey Telescope, Perth, Australia}
\affil{$^{25}$Solar-Terrestrial Environment Laboratory, Nagoya University, Nagoya, 464-8601, Japan}
\affil{$^{26}$Department of Physics, University of Auckland, Private Bag 92019, Auckland, New Zealand}
\affil{$^{27}$Department of Physics, University of Notre Damey, Notre Dame, IN 46556, USA}
\affil{$^{28}$Institute of Information and Mathematical Sciences, Massey University, Private Bag 102-904, North Shore Mail Centre, Auckland, New Zealand} 
\affil{$^{29}$Okayama Astrophysical Observatory, NAOJ, Okayama 719-0232, Japan}
\affil{$^{30}$University of Canterbury, Department of Physics and Astronomy, Private Bag 4800, Christchurch 8020, New Zealand}  
\affil{$^{31}$Mt. John Observatory, P.O. Box 56, Lake Tekapo 8770, New Zealand} 
\affil{$^{32}$School of Chemical and Physical Sciences, Victoria University, Wellington, New Zealand} 
\affil{$^{33}$Department of Physics, Konan University, Nishiokamoto 8-9-1, Kobe 658-8501, Japan} 
\affil{$^{34}$Nagano National College of Technology, Nagano 381-8550, Japan} 
\affil{$^{35}$Tokyo Metropolitan College of Industrial Technology, Tokyo 116-8523, Japan} 
\affil{$^{36}$Universidad de Concepci\'on, Departamento de Fisica, Casilla 160-C, Concepci{\'o}n, Chile} 
\affil{$^{37}$Institute of Astronomy Cambridge University, Madingley Road, CB3 0HA Cambridge, UK} 
\affil{$^{38}$LATT, Universit\'e de Toulouse, CNRS, 14 Avenue Edouard Belin, 31400 Toulouse, France} 
\affil{$^{39}$NASA Goddard Space Flight Center, 8800 Greenbelt Road, Greenbelt, MD 20771, USA}
\affil{$^{40}$Department of Physics, Massachusetts Institute of Technology, 77 Mass. Ave., Cambridge, MA 02139, USA}
\affil{$^{41}$European Southern Observatory, Casilla 19001, Vitacura 19, Santiago, Chile} 
\affil{$^{42}$School of Math and Physics, University of Tasmania, Private Bag 37, GPO Hobart, Tasmania 7001, Australia} 
\affil{$^{43}$Physics Department, Faculty of Arts and Sciences, University of Rijeka, Omladinska 14, 51000 Rijeka, Croatia}
\affil{$^{44}$Technical University of Vienna, Department of Computing, Wiedner Hauptstrasse 10, Vienna, Austria}
\affil{$^{45}$NASA Exoplanet Science Institute, Caltech, MS 100-22, 770 South Wilson Avenue, Pasadena, CA 91125, USA}
\affil{$^{46}$South African Astronomical Observatory, P.O. Box 9 Observatory 7935, South Africa} 
\affil{$^{47}$Space Telescope Science Institute, 3700 San Martin Drive, Baltimore, MD 21218, USA}
\affil{$^{48}$Astronomisches Rechen-Institut (ARI), Zentrum f{\"u}r Astronomie der Universit{\"a}t Heidelberg (ZAH), M{\"o}nchhofstrasse 12-14, 69120 Heidelberg, Germany}
\affil{$^{49}$Perth Observatory, Walnut Road, Bickley, Perth 6076, Australia}
\affil{$^{50}$School of Physics, University of Exeter, Stocker Road, Exeter, Devon, EX4 4QL, UK}
\affil{$^{51}$European Southern Observatory, Karl-Schwarzschild-Stra{\ss}e 2, 85748 Garching bei M{\"u}nchen, Germany}
\affil{$^{52}$Astrophysics Research Institute, Liverpool John Moores University, Egerton Wharf, Birkenhead CH41 1LD, UK}
\affil{$^{53}$Max-Planck-Institut f{\"o}r Sonnensystemforschung, Max-Planck-Str. 2, 37191 Katlenburg-Lindau, Germany}
\affil{$^{54}$Qatar Foundation, P.O. Box 5825, Doha, Qatar}
\affil{$^{55}$Department of Physics, University of Salerno, Via Ponte Don Melillo, 84084 Fisciano (SA), Italy}
\affil{$^{56}$Deutsches SOFIA Institut, Universit\"{a}t Stuttgart, Pfaffenwaldring 31, 70569 Stuttgart, Germany}
\affil{$^{57}$Institut f\"{u}r Astrophysik, Georg-August-Universit\"{a}t, Friedrich-Hund-Platz 1, 37077 G\"{o}ttingen, Germany}
\affil{$^{58}$Institut d'Astrophysique et de G\'{e}ophysique, All\'{e}e du 6 Ao\^{u}t 17, Sart Tilman, B\^{a}t.\ B5c, 4000 Li\`{e}ge, Belgium}
\affil{$^{59}$Department of Physics \& Astronomy, Aarhus University, Ny Munkegade 120, 8000 {\AA}rhus C, Denmark}
\affil{$^{60}$Niels Bohr Institutet, K{\o}benhavns Universitet, Juliane Maries Vej 30, 2100 K{\o}benhavn {\O}, Denmark}
\affil{$^{61}$Centre for Star and Planet Formation, Geological Museum, {\O}ster Voldgade 5, 1350 Copenhagen, Denmark}
\affil{$^{62}$Jodrell Bank Centre for Astrophysics, University of Manchester, Oxford Road,Manchester, M13 9PL, UK}
\affil{$^{63}$Max Planck Institute for Astronomy, K{\"o}nigstuhl 17, 69117 Heidelberg, Germany}
\affil{$^{64}$Department of Physics, Sharif University of Technology, P.O.~Box 11365--9161, Tehran, Iran}
\affil{$^{65}$Perimeter Institute for Theoretical Physics, 31 Caroline Street North, Waterloo, Ontario N2L 2Y5, Canada}
\affil{$^{66}$Astrophysics Group, Keele University, Staffordshire, ST5 5BG, UK}
\affil{$^{67}$Instituto Nacional de Pesquisas Espaciais/MCTI, S\~ao Jos\'e dos Campos, S\~ao Paulo, Brazil}
\affil{$^{68}$The $\mu$FUN Collaboration}
\affil{$^{69}$The MOA Collaboration}
\affil{$^{70}$The OGLE Collaboration}
\affil{$^{71}$The PLANET Collaboration}
\affil{$^{72}$The RoboNet Collaboration}
\affil{$^{73}$The MiNDSTEp Consortium}
\affil{$^{74}$Corresponding author}

\begin{abstract}
We present the analysis of the light curves of 9 high-magnification single-lens 
gravitational microlensing events with lenses passing over source stars, 
including OGLE-2004-BLG-254, MOA-2007-BLG-176, MOA-2007-BLG-233/OGLE-2007-BLG-302, 
MOA-2009-BLG-174, MOA-2010-BLG-436, MOA-2011-BLG-093, MOA-2011-BLG-274, 
OGLE-2011-BLG-0990/MOA-2011-BLG-300, and OGLE-2011-BLG-1101/MOA-2011-BLG-325.
For all events, we measure the linear limb-darkening coefficients of the 
surface brightness profile of source stars by measuring the deviation of 
the light curves near the peak affected by the finite-source effect.  For 
7 events, we measure the Einstein radii and the lens-source relative proper 
motions.  Among them, 5 events 
are found to have Einstein radii less than 0.2 mas, making the lenses candidates 
of very low-mass stars or brown dwarfs. For MOA-2011-BLG-274, especially, 
the small Einstein radius of $\theta_{\rm E}\sim 0.08$ mas combined with 
the short time scale of $t_{\rm E}\sim 2.7$ days suggests the possibility 
that the lens is a free-floating planet.  For MOA-2009-BLG-174, we measure 
the lens parallax and thus uniquely determine the physical parameters of the lens. 
We also find that the measured lens mass of $\sim 0.84\ M_\odot$ is consistent with 
that of a star blended with the source, suggesting that the blend is 
likely to be the lens. Although we find planetary signals for none of events,
we provide exclusion diagrams showing the confidence levels excluding the existence 
of a planet as a function of the separation and mass ratio.
\end{abstract}

\keywords{gravitational lensing: micro -- Galaxy: bulge}

\section{Introduction}

\begin{figure}[ht]
\epsscale{1.1}
\plotone{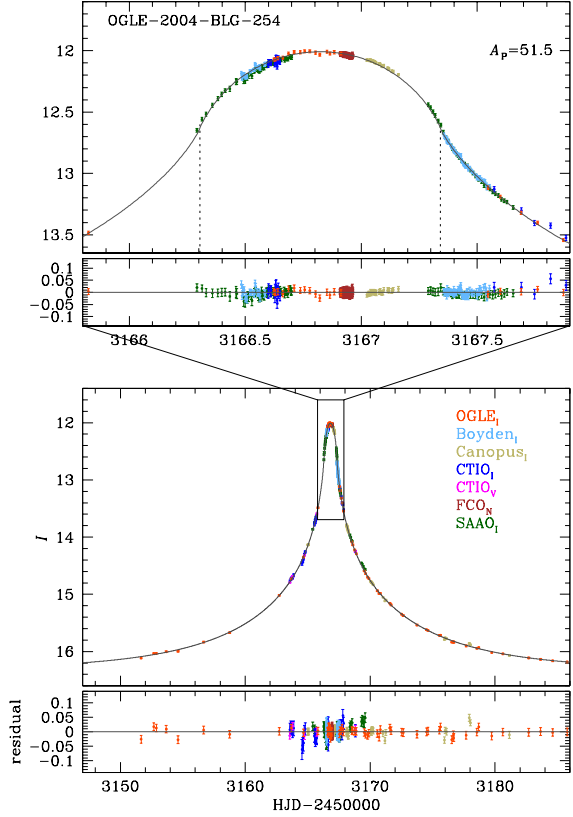}
\caption{\label{fig:one}
Light curve of OGLE-2004-BLG-254.  
The lower two panels show the overall shape of the light curve and residual 
from the best-fit model. The upper two panels show the enlargement 
of the peak region enclosed by a small box in the lower panel. 
We note that a model light curve varies depending on an observed passband 
due to the chromaticity caused by the finite-source effect. The presented
model curve is based on the passband of the first observatory in the list. 
However, the residuals of the individual data sets are based on the model 
curves of the corresponding passbands.
Colors of data points are chosen to match those of the labels of observatories 
where data were taken. 
The two dotted vertical lines in the upper panel represent the limb-crossing 
start/end times. 
The peak source magnification $A_{\rm P}$ is given in the upper panel.
}\end{figure}

\begin{figure}[ht]
\epsscale{1.1}
\plotone{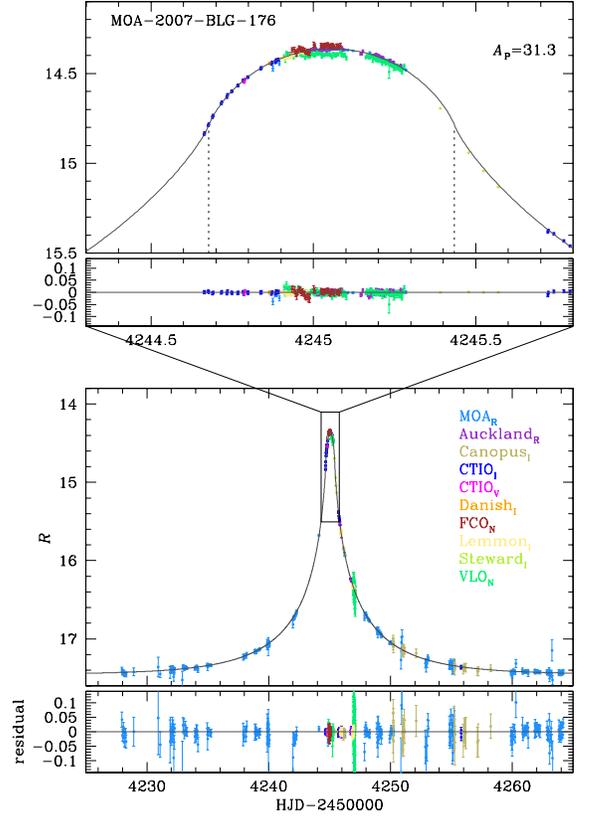}
\caption{\label{fig:two}
Light curve of MOA-2007-BLG-176. Notations are same as in Fig.1 
}\end{figure}

When an astronomical object (lens) is closely aligned with a background star
(source), the light from the source is deflected by the gravity of the lens,
resulting in brightening of the source star. The magnification of the source
flux is related to the projected lens-source separation by
\begin{equation}
A={u^2+2 \over u\sqrt{u^2+4}},
\label{eq1}
\end{equation}
where the separation $u$ is expressed in units of the angular Einstein radius 
$\theta_{\rm E}$.  The Einstein radius is related to the physical parameters 
of the lens system by
\begin{equation}
\theta_{\rm E}=\left( \kappa M \pi_{\rm rel}\right)^{1/2};\qquad
\pi_{\rm rel}={\rm AU} \left( {1\over D_{\rm L}}-{1\over D_{\rm S}}\right),
\label{eq2}
\end{equation}
where $\kappa=4G/(c^2{\rm AU})=8.14$ mas ${M_{\odot}}^{-1}$, $M$ is the mass of 
the lens, $\pi_{\rm rel}$ is the relative source-lens parallax, and 
$D_{\rm L}$ and $D_{\rm S}$ are the distances to the lens and 
source star, respectively. The relative motion between the source, lens, and 
observer leads to light variation of the source star (lensing event). 
The first microlensing events were detected by \citet{alcock93}  and 
\citet{udalski93} from the experiments conducted based on the proposal of 
\citet{paczynski86}. With the development of observational strategy combined 
with upgraded instrument, the detection rate of lensing events has been 
dramatically increased from several dozen events per year during 
the early phase of lensing experiments to more than a thousand events 
per year in current experiments.

The magnification of source star flux increases as the lens approaches closer
to the source star. For a small fraction of events, the lens-source separation 
is even smaller than the source radius and the lens passes over the surface of 
the source star. These events are of scientific importance 
due to various reasons.

First, a high-magnification event with a lens passing over a source star provides 
a rare chance to measure the brightness profile of a remote star. For such an 
event, in which the lens-source separation is comparable to the source size near 
the peak of the event, different parts of the source star are magnified 
by different amounts. The resulting lensing light curve 
deviates from the standard form of a point-source event \citep{witt94, gould94, 
nemiroff94, alcock97} and the analysis of the deviation enables to measure 
the limb-darkening profile of the lensed star \citep{witt95, loeb95, valls98, 
bryce02, heyrovsky03}. With the same principle, it is also possible to study 
irregular surface structures such as spots \citep{heyrovsky00, han00, hendry02, 
rattenbury02}.

\begin{deluxetable*}{ll}
\tablecaption{Events With Lenses Passing Over Source Stars\label{table:one}}
\tablewidth{0pt}
\tablehead{
\multicolumn{1}{c}{event} &
\multicolumn{1}{c}{reference}
}
\startdata
OGLE-2004-BLG-254                   & \citet{cassan06} / this work  \\
OGLE-2004-BLG-482                   & \citet{zub11}                 \\
MOA-2006-BLG-130/OGLE-2006-BLG-437  & \citet{baudry11} / under analysis  \\
OGLE-2007-BLG-050/MOA-2007-BLG-103  & \citet{batista09}             \\
MOA-2007-BLG-176                    & this work                     \\
OGLE-2007-BLG-224/MOA-2007-BLG-163  & \citet{gould09}               \\
MOA-2007-BLG-233/OGLE-2007-BLG-302  & this work                     \\
OGLE-2008-BLG-279/MOA-2008-BLG-225  & \citet{yee09}                 \\
OGLE-2008-BLG-290/MOA-2008-BLG-241  & \citet{fouque10}              \\
MOA-2009-BLG-174                    & this work                     \\
MOA-2009-BLG-411                    & \citet{fouque11} / under analysis  \\
MOA-2010-BLG-311                    & \citet{hung11} / under analysis    \\
MOA-2010-BLG-436                    & this work                     \\
MOA-2010-BLG-523                    & \citet{gould11} / under analysis \\
MOA-2011-BLG-093                    & this work                     \\
MOA-2011-BLG-274                    & this work                     \\
OGLE-2011-BLG-0990/MOA-2011-BLG-300 & this work                     \\
OGLE-2011-BLG-1101/MOA-2011-BLG-325 & this work                    
\enddata  
\end{deluxetable*}

Second, it is possible to measure the Einstein radius of the lens and the 
relative lens-source proper motion. The light curve at the moment of the 
entrance (exit) of the lens into (from) the source surface exhibits inflection 
of the curvature.  The duration of the passage over the source as 
measured by the interval between the entrance and exit of the lens over the 
surface of the source star is
\begin{equation}
\Delta t_{\rm T} = 2\sqrt{{\rho_\star}^2-{u_0}^2}\ t_{\rm E},
\label{eq3}
\end{equation}
where $\rho_{\star}$ is the source radius in units of $\theta_{\rm E}$ 
(normalized source radius), $u_0$ is the lens-source separation normalized by 
$\theta_{\rm E}$ at the moment of the closest approach (impact parameter), 
and $t_{\rm E}$ is the time scale for the lens to transit $\theta_{\rm E}$ 
(Einstein time scale). The impact parameter and the Einstein time scale are 
measured from the overall shape of the light curve and the duration of the 
event. With the known $u_0$ and $t_{\rm E}$ combined with the measured 
duration of passage over the source, the normalized source radius is measured 
from the relation (\ref{eq3}). With the additional information about the angular 
source size, $\theta_{\star}$, then the Einstein radius and the lens-source 
proper motion are measured as $\theta_{\rm E}=\theta_\star/\rho_\star$ and 
$\mu= \theta_{\rm E}/t_{\rm E}$, respectively.  For general lensing events, 
the Einstein time scale is the only measurable quantity related to the physical 
parameters of the lens. However, the time scale results from the combination of 
3 physical parameters of the mass of the lens, $M$, the distance to the lens, 
$D_{\rm L}$, and the lens-source transverse speed, $v$, and thus the information 
about the lens is highly degenerate. The Einstein radius, on the other hand, 
does not depend on $v$ and thus the physical parameters of the lens can be 
better constrained. For a fraction of events with long time scales, it is 
possible to additionally measure the lens parallax, $\pi_{\rm E}=\pi_{\rm rel}
/\theta_{\rm E}$, from the deviation of the light curve induced by the orbital 
motion of the Earth around the Sun.  With the Einstein radius and the lens 
parallax measured, the physical parameters of the lens are uniquely determined 
\citep{gould97}.

Third, high-magnification events are sensitive to planetary companions of
lenses. This is because a planet induces a small caustic near the primary lens
and a high-magnification event resulting from the source trajectory passing
close to the primary has a high chance to produce signals indicating the
existence of the planet \citep{griest98}. For an event with a lens passing over 
a source star, the planetary signal is weakened by the finite-source effect 
\citep{bennett96}.  Nevertheless, two of the microlensing planets were discovered 
through this channel: MOA-2007-BLG-400 \citep{dong09} and 
MOA-2008-BLG-310 \citep{janczak10}.

Fourth, high-magnification events provide a chance to spectroscopically study
remote Galactic bulge stars. Most stars in the Galactic bulge are too faint 
for spectroscopic observations even with large telescopes. However, enhanced 
brightness of lensed stars of high-magnification events allows spectroscopic 
observation possible, enabling population study of Galactic bulge stars 
\citep{johnson08, bensby09, bensby11, cohen09, epstein10}.

In this work, we present integrated results of analysis for 14  
high-magnification events with lenses passing over source stars that have 
been detected since 2004. Among them, 8 events were newly analyzed and one 
event was reanalyzed with additional data.

\begin{deluxetable*}{llllll}
\tabletypesize{\small}
\tablecaption{Observatories\label{table:two}}
\tablewidth{0pt}
\tablehead{
\multicolumn{1}{c}{event} &
\multicolumn{1}{c}{MOA} &
\multicolumn{1}{c}{OGLE} &
\multicolumn{1}{c}{$\mu$FUN} &
\multicolumn{1}{c}{PLANET} &
\multicolumn{1}{c}{RoboNet} \\
\multicolumn{1}{c}{(RA,DEC)$_{J2000}$} &
\multicolumn{1}{c}{} &
\multicolumn{1}{c}{} &
\multicolumn{1}{c}{} &
\multicolumn{1}{c}{} &
\multicolumn{1}{c}{/MiNDSTEp} 
}
\startdata
OGLE-2004-BLG-254                                                                                       &                    & $\rm{LCO_I}$  & $\rm{CTIO_{I,V}}$ (39/5)   & $\rm{Boyden_I}$ (74)        &                                           \\
($17^{\rm h}56^{\rm m}36^{\rm s}\hskip-2pt.20$,$-32^{\circ}33^{\prime}01^{\prime\prime}\hskip-2pt.80$)  &                    & (377)         & $\rm{FCO_N}$ (129)         & $\rm{Canopus_I}$ (59)       &                                           \\
                                                                                                        &                    &               &                            & $\rm{SAAO_I}$ (112)         &                                           \\
\hline                                                                                                                                                                                                
MOA-2007-BLG-176                                                                                        & $\rm{Mt. John_R}$  &               & $\rm{Auckland_R}$ (68)     & $\rm{Canopus_I}$ (26)       &                     $\rm{Danish_I}$ (2)   \\
($18^{\rm h}05^{\rm m}00^{\rm s}\hskip-2pt.41$,$-25^{\circ}47^{\prime}03^{\prime\prime}\hskip-2pt.69$)  & (1388)             &               & $\rm{CTIO_{I,V}}$ (41/4)   & $\rm{Steward_I}$ (4)        &                                           \\
                                                                                                        &                    &               & $\rm{FCO_N}$ (33)          &                             &                                           \\
                                                                                                        &                    &               & $\rm{Lemmon_I}$ (66)       &                             &                                           \\
                                                                                                        &                    &               & $\rm{VLO_N}$ (129)         &                             &                                           \\
\hline                                                                                                                                                                   
MOA-2007-BLG-233                                                                                        & $\rm{Mt. John_R}$  & $\rm{LCO_I}$  & $\rm{CTIO_{I,V}}$ (80/5)   & $\rm{Canopus_{I,V}}$ (60/5) &                     $\rm{Danish_I}$ (125) \\
/OGLE-2007-BLG-302                                                                                      & (645)              & (628)         & $\rm{FCO_N}$ (23)          & $\rm{Perth_I}$ (23)         &                                           \\
($17^{\rm h}54^{\rm m}14^{\rm s}\hskip-2pt.86$,$-31^{\circ}11^{\prime}02^{\prime\prime}\hskip-2pt.65$)  &                    &               & $\rm{Lemmon_I}$ (19)       & $\rm{SAAO_I}$ (80)          &                                           \\
                                                                                                        &                    &               & $\rm{SSO_N}$ (80)          &                             &                                           \\
\hline                                                                                                                                                                   
MOA-2009-BLG-174                                                                                        & $\rm{Mt. John_R}$  &               & $\rm{Bronberg_N}$ (147)    & $\rm{Canopus_I}$ (40)       & $\rm{LT_R}$ (7)                           \\ 
($18^{\rm h}02^{\rm m}07^{\rm s}\hskip-2pt.60$,$-31^{\circ}25^{\prime}24^{\prime\prime}\hskip-2pt.20$)  & (2189)             &               & $\rm{CAO_N}$ (111)         &                             &                                           \\
                                                                                                        &                    &               & $\rm{Craigie_N}$ (130)     &                             &                                           \\   
                                                                                                        &                    &               & $\rm{CTIO_{I,V}}$ (286/7)  &                             &                                           \\   
                                                                                                        &                    &               & $\rm{Kumeu_N}$ (90)        &                             &                                           \\   
                                                                                                        &                    &               & $\rm{Possum_N}$ (60)       &                             &                                           \\   
\hline                                                                                                                                                                   
MOA-2010-BLG-436                                                                                        & $\rm{Mt. John_R}$  &               &                            & $\rm{SAAO_{I,V}}$ (14/3)    & $\rm{FTS_I}$ (3)                          \\
($18^{\rm h}03^{\rm m}21^{\rm s}\hskip-2pt.68$,$-27^{\circ}38^{\prime}10^{\prime\prime}\hskip-2pt.74$)  & (2581)             &               &                            &                             &                                           \\
\hline                                                                                                                                                                   
MOA-2011-BLG-093                                                                                        & $\rm{Mt. John_R}$  & $\rm{LCO_I}$  & $\rm{CTIO_{I,V}}$ (76/21)  & $\rm{Canopus_I}$ (254)      & $\rm{FTN_I}$ (3)                          \\
($17^{\rm h}46^{\rm m}17^{\rm s}\hskip-2pt.83$,$-34^{\circ}20^{\prime}24^{\prime\prime}\hskip-2pt.76$)  & (2247)             & (292)         & $\rm{PEST_N}$ (124)        &                             & $\rm{FTS_I}$ (19)                         \\   
\hline                                                                                                                                                                   
MOA-2011-BLG-274                                                                                        & $\rm{Mt. John_R}$  & $\rm{LCO_I}$  & $\rm{Auckland_R}$ (53)     &                             &                                           \\
($17^{\rm h}54^{\rm m}42^{\rm s}\hskip-2pt.34$,$-28^{\circ}54^{\prime}59^{\prime\prime}\hskip-2pt.26$)  & (3447)             & (76)          & $\rm{CTIO_I}$ (4)          &                             &                                           \\   
                                                                                                        &                    &               & $\rm{FCO_N}$ (16)          &                             &                                           \\   
                                                                                                        &                    &               & $\rm{Kumeu_R}$ (49)        &                             &                                           \\   
                                                                                                        &                    &               & $\rm{PEST_N}$ (15)         &                             &                                           \\   
\hline                                                                                                                                                                   
OGLE-2011-BLG-0990                                                                                      & $\rm{Mt. John_R}$  & $\rm{LCO_I}$  & $\rm{OPD_I}$ (275)         & $\rm{Canopus_I}$ (10)       &                                           \\
/MOA-2011-BLG-300                                                                                       & (1708)             & (3434)        & $\rm{Possum_R}$ (23)       & $\rm{SAAO_{I,V}}$ (95/6)    &                                           \\
($17^{\rm h}51^{\rm m}30^{\rm s}\hskip-2pt.29$,$-30^{\circ}17^{\prime}47^{\prime\prime}\hskip-2pt.60$)  &                    &               &                            &                             &                                           \\
\hline                                                                                                     
OGLE-2011-BLG-1101                                                                                      & $\rm{Mt. John_R}$  & $\rm{LCO_I}$  & $\rm{Auckland_R}$ (60)     & $\rm{Canopus_I}$ (98)       & $\rm{FTN_I}$ (65)                         \\
/MOA-2011-BLG-325                                                                                       & (609)              & (192)         & $\rm{CTIO_{I,V}}$ (126/12) &                             & $\rm{FTS_I}$ (145)                        \\
($18^{\rm h}03^{\rm m}31^{\rm s}\hskip-2pt.62$,$-26^{\circ}20^{\prime}39^{\prime\prime}\hskip-2pt.50$)  &                    &               & $\rm{Possum_R}$ (24)       &                             & $\rm{LT_I}$ (27)                          \\
                                                                                                        &                    &               & $\rm{SSO_N}$ (107)         &                             &                                           \\
                                                                                                        &                    &               & $\rm{VLO_N}$ (113)         &                             &                                             
\enddata  
\tablecomments{ 
Mt. John: Mt. John Observatory, New Zealand; LCO: Las Campanas Observatory, 
Chile; Auckland: Auckland Observatory, New Zealand; Bronberg: Bronberg 
Observatory, South Africa; CAO: CAO San Pedro Observatory, Chile; Cragie: 
Craigie Observatory, Australia; CTIO: Cerro Tololo Inter-American Observatory, 
Chile; FCO: Farm Cove Observatory, New Zealand; Kumeu: Kumeu Observatory, New 
Zealand; Lemmon: Mt Lemmon Observatory, Arizona, USA; OPD: Observatorio do 
Pico dos Dias, Brazil; PEST: Perth Exoplanet Survey Telescope, Australia; 
Possum: Possum Observatory, New Zealand; SSO: Southern Stars Observatory, 
French Polynesia; VLO: Vintage Lane Observatory, New Zealand; 
Boyden: Boyden Observatory, South Africa; Canopus: Canopus Hill Observatory, 
Tasmania, Australia; Perth: Perth Observatory, Australia; SAAO: South African 
Astronomical Observatory, South Africa; Steward: Steward Observatory, Arizona, 
USA; FTN: Faulkes North, Hawaii; FTS: Faulkes South, Australia; LT: Liverpool 
Telescope, La Palma, Spain; Danish: Danish Telescope, European Southern Observatory, 
La Silla, Chile. The subscription after each observatory represents the filter 
used for observation and the value in parenthesis is the number of data points. 
The filter ``N'' denotes that no filter is used.
}
\end{deluxetable*}

\section{Event Selection}

The sample of events in our analysis is selected under the definition of {\it a single-lens
event where the lens-source separation at the time of the peak magnification 
is less than the radius of the source star, i.e. $u_0 < \rho_*$ and thus 
the lens passes over the surface of the source star.} To obtain a sample of events, we 
begin with searching for high-magnification events that have been detected since 
2004. Events with lenses passing over source stars can be usually distinguished 
by the characteristic features of their light curves near the peak. These features 
are the inflection of the curvature at the moment when the finite source 
first touches and completely leaves the lens and the round shape of the light 
curve during the passage of the lens over the source. To be more objective 
than visual inspection, we conduct modeling of all high-magnification events 
with peak magnifications $A_{\rm P}\ge10$ to judge the qualification of events. 
From these searches, we find that there exist 18 such events. Among them, 
analysis results of 12 events were not published before. We learn that 
4 unpublished events MOA-2006-BLG-130/OGLE-2006-BLG-437 \citep{baudry11}, 
MOA-2009-BLG-411 \citep{fouque11}, MOA-2010-BLG-523 \citep{gould11}, 
and MOA-2010-BLG-311 \citep{hung11} are under analysis by other researchers 
and thus exclude them in our analysis. We note that there exist 
4 known source-crossing events detected before 2004, including
MACHO Alert 95-30 \citep{alcock97}, OGLE sc26\_2218 \citep{smith03}, 
OGLE-2003-BLG-238 \citep{jiang04}, and OGLE-2003-BLG-262 \citep{yoo04}. We also 
note that MOA-2007-BLG-400 \citep{dong09} and MOA-2008-BLG-310 \citep{janczak10} 
exhibit characteristic features of source-crossing single-lens events but 
we exclude them in the sample because the lenses of the events turned out to 
have planetary companions.

In this work, we conduct analyses of 9 events.  Among them, 8 events are 
newly analyzed in this work. These events include MOA-2007-BLG-176, 
MOA-2007-BLG-233/OGLE-2007-BLG-302, MOA-2009-BLG-174, MOA-2010-BLG-436, 
MOA-2011-BLG-093, MOA-2011-BLG-274, OGLE-2011-BLG-0990/MOA-2011-BLG-300, 
and OGLE-2011-BLG-1101/MOA-BLG-2011-325. For OGLE-2004-BLG-254, which was 
analyzed before by \citet{cassan06}, we conduct additional analysis by 
adding more data sets taken from CTIO and FCO. \footnote{Besides the data 
sets listed in Table \ref{table:two}, there exists an additional data set taken 
by using the Danish telescope. However, we do not use these data because 
it has been shown by \citet{heyrovsky08} that the large scatter of the 
data results in poor measurement of lensing parameters including the 
limb-darkening coefficient.} In Table \ref{table:one}, we summarize 
the status of analysis for the total 18 events that have been detected since 2004.

\section{Observation}

For almost all events analyzed in this work, the source-crossing part of 
the light curve was densely covered. This was possible due to the 
coordinated work of survey and follow-up observations. Survey groups issued 
alerts of events. For a fraction of the events with high-magnifications,
additional alerts were issued.  In other cases, follow-up teams issued 
high-magnification alerts independently.  The peak time of a high-magnification 
event was predicted by real-time modeling based on the rising part of 
the light curve.  Finally, the peak was densely covered by many telescopes 
that were prepared for follow-up observations at the predicted time of the peak.  
For MOA-2010-BLG-436, the rising part of the light curve was not covered by 
survey observations due to the short time scale of the event and thus 
no alert was issued. Nevertheless, the event was positioned in a high frequency 
field of the MOA survey and thus the peak was covered densely enough to be 
confirmed as an event with the lens passing over the source.

\begin{figure}[ht]
\epsscale{1.1}
\plotone{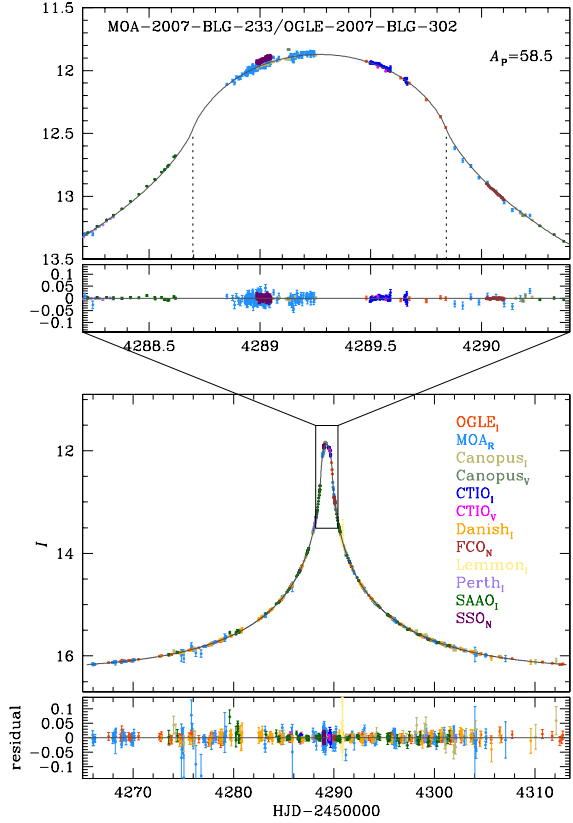}
\caption{\label{fig:three}
Light curve of MOA-2007-BLG-233/OGLE-2007-BLG-302. Notations are same as in 
Fig.1 
}\end{figure}

\begin{figure}[ht]
\epsscale{1.1}
\plotone{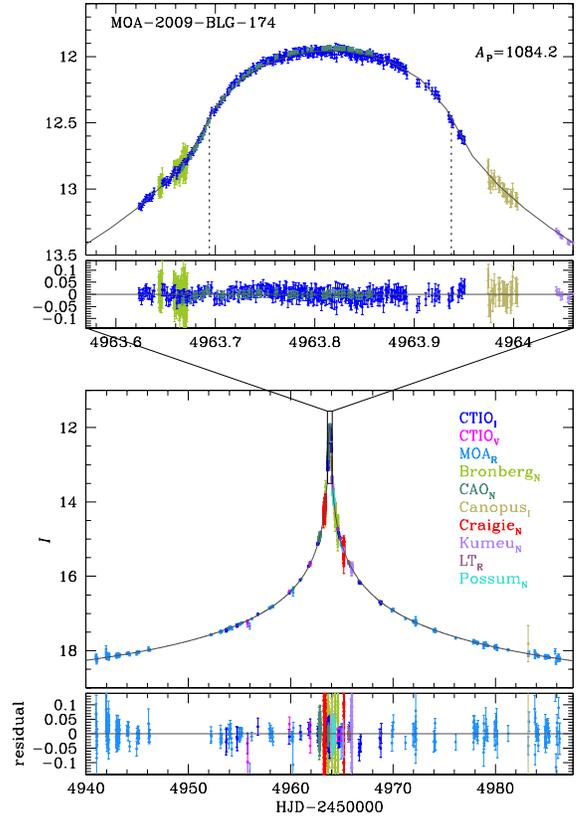}
\caption{\label{fig:four}
Light curve of MOA-2009-BLG-174. Notations are same as in Fig.1  
}\end{figure}

Table \ref{table:two} shows the observatories of the telescopes that were 
used for observations of the individual events along with the observed passbands 
(marked as subscripts after the observatory names) and the numbers of data 
points (values in parentheses). Also marked are the coordinates (RA,DEC) of 
the events. Survey observations were conducted by MOA 
and OGLE groups using the 1.8 m telescope at Mt. John Observatory in New 
Zealand and the 1.3m Warsaw University telescope at Las Campanas Observatory 
in Chile, respectively. Follow-up observations were carried out by $\mu$FUN, 
PLANET, RoboNet, and MiNDSTEp groups using 22 telescopes located in 8 different 
countries. These telescopes include 1.3 m SMARTS CTIO, 0.4 m CAO in Chile, 
0.4 m Auckland, 0.4 m FCO, 0.4 m Possum, 0.4 m Kumeu, 0.4 m VLO in New Zealand, 
1.0 m Lemmon in Arizona, USA, 0.4 m Bronberg in South Africa, 0.6 m Pico dos Dias in Brazil, 
0.25 m Craigie, 0.3 m PEST in Australia, 0.28m SSO in French Polynesia, 1.0 m SAAO, 
1.5 m Boyden in South Africa, 1.0 m Canopus, 0.6 m Perth in Australia, 
1.5 m Steward in Arizona, USA, 2.0 m FTN in Hawaii, USA, 2.0 m FTS in Australia, 
2.0 m LT in La Palma, Spain, and 1.54 m Danish in La Silla, Chile.

Reduction of data was conducted by using photometry codes that were developed 
by the individual groups. The MOA and OGLE data were reduced by photometry 
codes developed by \citet{bond01} and \citet{udalski03}, respectively, which are 
based on Difference Image Analysis method \citep{alard98}. The $\mu$FUN data were 
processed using a DoPHOT pipeline \citep{schechter93}.  For PLANET and MiNDSTEp data, 
a pySIS pipeline \citep{albrow09} is used. For RoboNet data, a DanDIA pipeline 
\citep{bramich08} is used.

The error bars estimated from different observatories are rescaled so that 
$\chi^2$ per degree of freedom becomes unity for the data set of each observatory 
where $\chi^2$ is computed based on the best-fit model. According to this simple 
scheme, however, we find a systematic tendency for some data sets that error bars near the 
peak of a light curve are overestimated. We find that this is caused by the inclusion 
of redundant data at the baseline in error normalization. In this case, the data at the baseline 
greatly outnumber accurate data points near the peak and thus error-bar normalization 
is mostly dominated by the baseline data. To minimize this systematics, we restrict 
the range of data for error normalization not to be too wide so that error estimation 
is not dominated by data at the baseline, but not to be too narrow so that lensing 
parameters can be measured accurately. For the final data set used for modeling, 
we eliminate data points lying beyond 3$\sigma$ from the best-fit model.

\begin{figure}[ht]
\epsscale{1.1}
\plotone{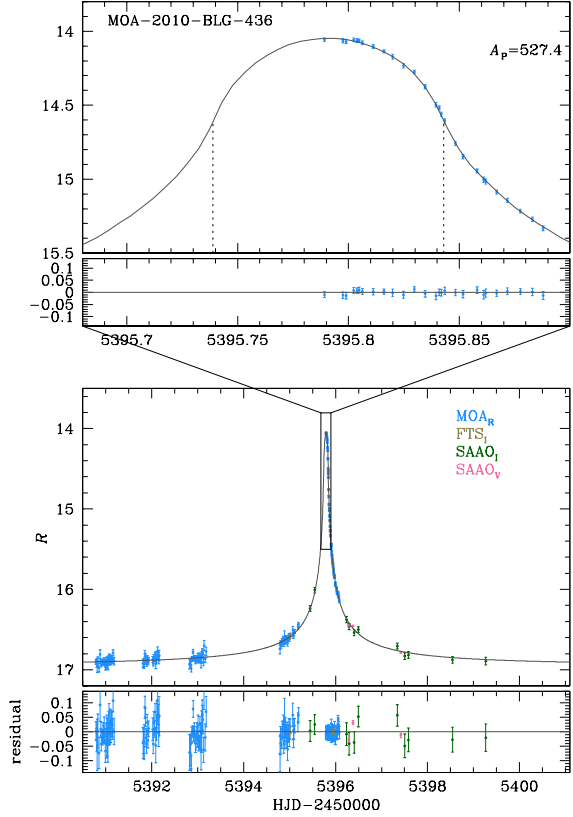}
\caption{\label{fig:five}
Light curve of MOA-2010-BLG-436. Notations are same as in Fig.1  
}\end{figure}

\begin{figure}[ht]
\epsscale{1.1}
\plotone{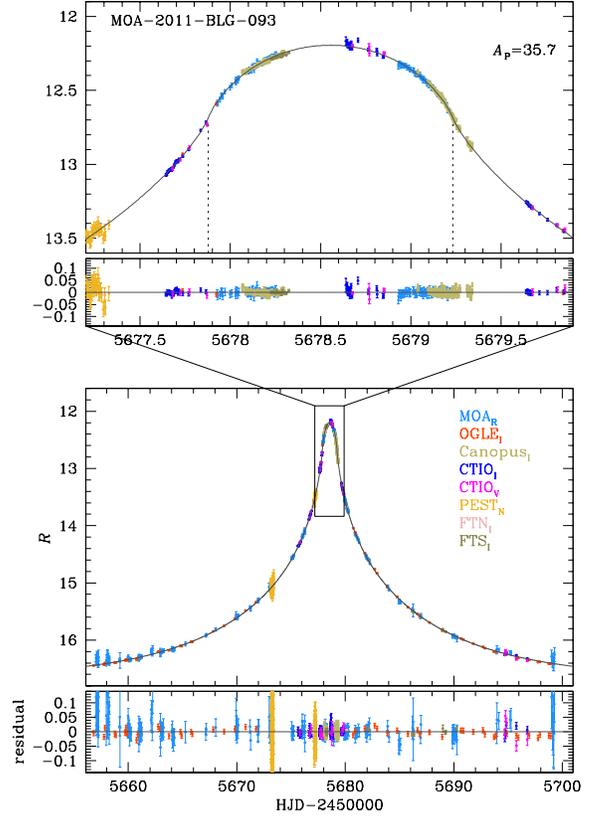}
\caption{\label{fig:six}
Light curve of MOA-2011-BLG-093. Notations are same as in Fig.1  
}\end{figure}

\begin{figure}[ht]
\epsscale{1.1}
\plotone{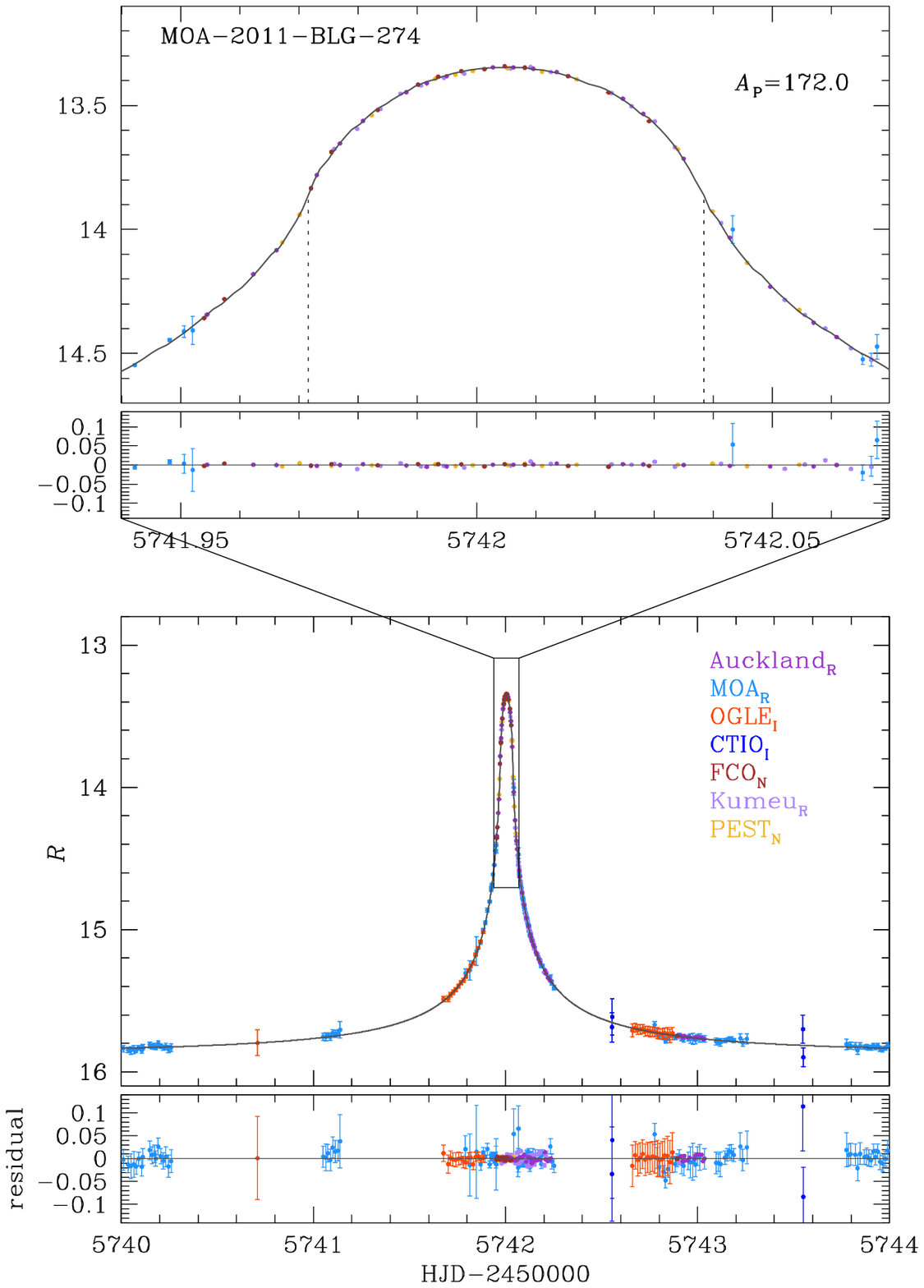}
\caption{\label{fig:seven}
Light curve of MOA-2011-BLG-274. Notations are same as in Fig.1  
}\end{figure}

\begin{figure}[ht]
\epsscale{1.1}
\plotone{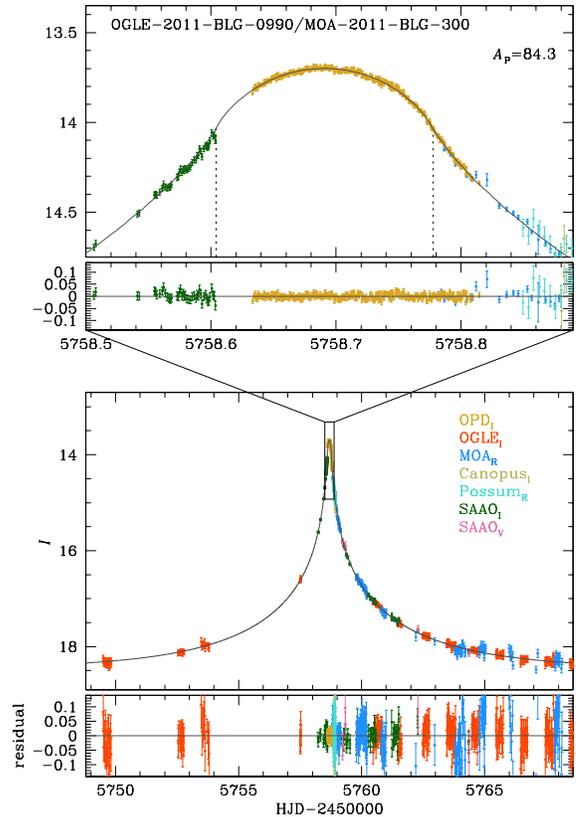}
\caption{\label{fig:eight}
Light curve of OGLE-2011-BLG-0990/MOA-2011-BLG-300. Notations are same as in Fig.1
}\end{figure}

\begin{figure}[ht]
\epsscale{1.1}
\plotone{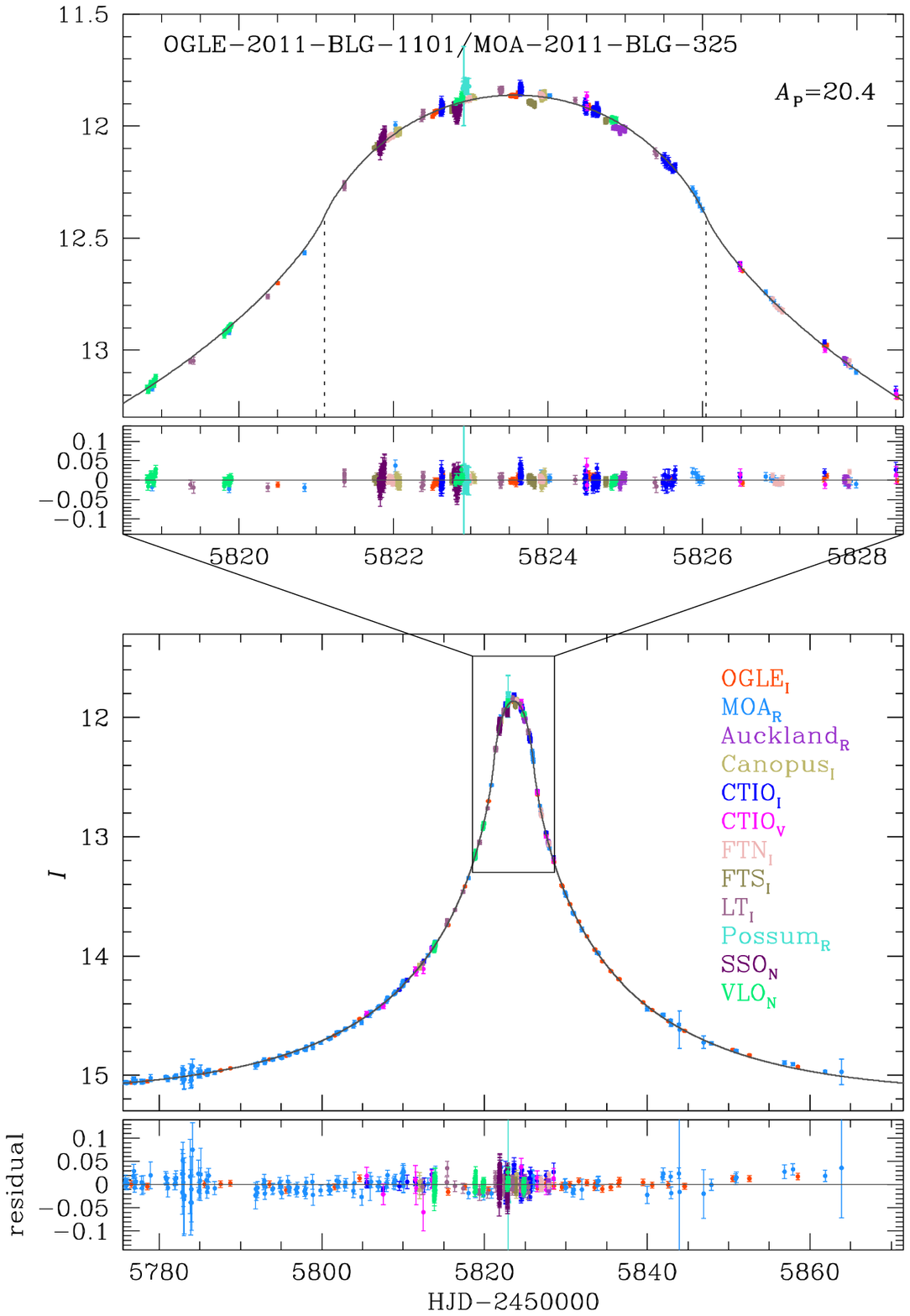}
\caption{\label{fig:nine}
Light curve of OGLE-2011-BLG-1101/MOA-2011-BLG-325. Notations are same as in Fig.1
}\end{figure}

In Figure \ref{fig:one}-\ref{fig:nine}, we present the light curves of the 
individual events. In each figure, the lower two panels show the overall 
shape of the light curve and residual and the upper two panels show the 
enlargement of the peak region of the light curve and residual. For each figure, 
we mark the moments of the lens'entrance and exit of the source by two dotted 
vertical lines. Also marked is the peak source magnification. We note that 
the same color of data points is used for each observatory throughout the light 
curves and colors of data points are chosen to match those of labels of 
observatories. We note that the magnitude scale corresponds to one of the observatories 
in the list, while data from the other observatories have adjusted blends and are 
vertically shifted to match the first light curve. The choice of reference is 
based on data from survey observation, i.e. OGLE and MOA data. If both OGLE 
and MOA data are available, the OGLE data is used for reference.

\section{Modeling}

Modeling the light curve of each event is conducted by searching for a set 
of lensing parameters that best describes the observed light curve. 
For all events, the light curves appear to have a standard form except 
the peak region and thus we start with a simple single-lens modeling. 
The light curve of a standard single-lensing event is characterized by 
3 parameters, that are the time of the closest lens-source approach, $t_0$, 
the lens-source separation at that moment, $u_0$, and the Einstein time scale, 
$t_{\rm E}$. Based on the initial solution, we refine the solution by 
considering additional second-order effects.

To precisely describe the peak region of the light curve of an event with 
a lens passing over a source, additional parameters are needed to describe 
the deviation caused by the finite-source effect. To the first order 
approximation, the finite-source effect is described by the normalized 
source radius, $\rho_\star$. For more refined description of the deviation, 
additional parameters of the limb-darkening coefficients, $u_\lambda$, 
are needed to account for the variation of the deviation caused by 
the brightness profile of the source star surface. With the coefficients, 
the limb-darkening profile is modeled by the standard linear law
\begin{equation}
I=I_0{\left[1-u_\lambda\left(1-cos\phi\right)\right]},
\label{eq4}
\end{equation}
where $I_0$ is the intensity of light at the center of the stellar 
disk and $\phi$ is the angle between the normal to the stellar surface and 
the line of sight toward the observer.

For an event with a time scale comparable to the orbital period of the Earth, 
the position of the observer changes by the Earth's orbital motion during 
the event and the resulting light curve deviates from a symmetric standard form. 
This parallax effect is described by 2 parameters of $\pi_{{\rm E},N}$ and 
$\pi_{{\rm E},E}$, that represent the two components of the lens parallax 
vector $\pivec_{\rm E}$ projected on the sky in the north and east equatorial 
coordinates, respectively. The direction of the parallax vector corresponds to the 
lens-source relative motion in the frame of the Earth. The size of the 
parallax vector corresponds to the ratio of the Earth's orbit, i.e. 1 AU, 
to the Einstein radius projected on the observer's plane.

For a high-magnification event, the peak of the light curve can exhibit 
additional deviations if the lens has a companion. For a planetary companion 
located near the Einstein ring of the primary lens or a binary companion 
with a separation from the primary substantially smaller or larger than 
the Einstein radius, a small caustic is induced near the location of the 
primary lens. Then, the source trajectory of a high-magnification event 
passing close to the primary lens has a high chance to approach the caustic, 
resulting in a perturbation near the peak of the light curve.  Description 
of the perturbation induced by a lens companion requires 3 additional 
parameters of the mass ratio between the binary lens components, $q$, the 
projected separation in units of the Einstein radius, $s$, and the angle 
between the source trajectory and the binary axis, $\alpha$.

For each event, we search for a solution of the best-fit lensing parameters by 
minimizing $\chi^2$ in the parameter space. For the $\chi^2$ minimization, 
we use a Markov Chain Monte Carlo (MCMC) method. We compute finite 
magnifications by using the ray-shooting technique \citep{schneider86, 
kayser86, wambsganss97}.  In this method, rays are uniformly shot from the 
image plane, bent according to the lens equation, and land on the source 
plane. Then, a finite magnification is computed by comparing the number 
densities of rays on the image and source planes. Precise computation of 
finite magnifications by using this numerical technique requires a large 
number of rays and thus demands heavy computation.  To minimize computation, 
we limit finite-magnification computation by using the ray-shooting method 
only when the lens is close to the source. Once a solution of the parameters 
is found, we estimate the uncertainties of the individual parameters based on 
the chain of solutions obtained from MCMC runs.

\begin{deluxetable*}{lllllll}
\tablecaption{Best-fit Parameters\label{table:three}}
\tablewidth{0pt}
\tablehead{
\multicolumn{1}{c}{event} &
\multicolumn{1}{c}{$\chi^{2}/{\rm dof}$} &
\multicolumn{1}{c}{$t_0$ (HJD-2450000)} &
\multicolumn{1}{c}{$u_0$} &
\multicolumn{1}{c}{$t_{\rm E}$ (days)} &
\multicolumn{1}{c}{$\rho_\star$} &
\multicolumn{1}{c}{$\pi_{\rm E}$}
}
\startdata
OGLE-2004-BLG-254       & 1326/593     & 3166.8194     &  0.0046         & 13.23       & 0.0400        & --             \\
                        &              & $\pm$0.0002   &  $\pm$0.0008    & $\pm$0.05   & $\pm$0.0002   & --             \\
OGLE-2004-BLG-482       & 756.9/693    & 3235.7816     &  0.000          & 9.61        & 0.1309        & --             \\
                        &              & $\pm$0.0007   &  $\pm$0.002     & $\pm$0.02   & $\pm$0.0005   & --             \\
OGLE-2007-BLG-050       & 1760.5/1745  & 4221.9726     &  0.002          & 68.09       & 0.0045        & 0.12           \\
/MOA-2007-BLG-103       &              & $\pm$0.0001   &  $\pm$0.000     & $\pm$0.66   & $\pm$0.0001   & $\pm$0.03      \\
OGLE-2007-BLG-224       & --           & --            &  0.00029        & 6.91        & 0.0009        & 1.97           \\
/MOA-2007-BLG-163       &              & --            &  --             & $\pm$0.13   & $\pm$0.0002   & $\pm$0.13      \\
OGLE-2008-BLG-279       & --           & 4617.34787    &  0.00066        & 106.0       & 0.00068       & 0.15           \\
/MOA-2008-BLG-225       &              & $\pm$0.00008  &  $\pm$0.00005   & $\pm$0.9    & $\pm$0.00006  & $\pm$0.02      \\
OGLE-2008-BLG-290       & 2317.7/2015  & 4632.56037    &  0.00276        & 16.36       & 0.0220        & --             \\
/MOA-2008-BLG-241       &              & $\pm$0.00027  &  $\pm$0.0002    & $\pm$0.08   & $\pm$0.0001   & --             \\
\hline
OGLE-2004-BLG-254       & 785.2/784    & 3166.823      &  0.0111         & 12.84       & 0.0418        & --             \\
                        &              & $\pm$0.001    &  $\pm$0.0004    & $\pm$0.09   & $\pm$0.0004   & --             \\
MOA-2007-BLG-176        & 1756.0/1747  & 4245.056      &  0.0363         & 8.13        & 0.0590        & --             \\
                        &              & $\pm$0.001    &  $\pm$0.0005    & $\pm$0.07   & $\pm$0.0006   & --             \\
MOA-2007-BLG-233        & 1779.4/1757  & 4289.269      &  0.0060         & 15.90       & 0.0364        & --             \\
/OGLE-2007-BLG-302      &              & $\pm$0.001    &  $\pm$0.0002    & $\pm$0.05   & $\pm$0.0001   & --             \\
MOA-2009-BLG-174        & 2816.5/3051  & 4963.816      &  0.0005         & 64.99       & 0.0020        & 0.06           \\ 
                        &              & $\pm$0.001    &  $\pm$0.0001    & $\pm$0.61   & $\pm$0.0001   & +0.07-0.02     \\ 
MOA-2010-BLG-436        & 2599.4/2593  & 5395.791      &  0.0002         & 12.78       & 0.0041        & --             \\
                        &              & $\pm$0.001    &  $\pm$0.0002    & $\pm$1.08   & $\pm$0.0003   & --             \\
MOA-2011-BLG-093        & 3038.0/3024  & 5678.555      &  0.0292         & 14.97       & 0.0538        & --             \\
                        &              & $\pm$0.001    &  $\pm$0.0002    & $\pm$0.05   & $\pm$0.0002   & --             \\
MOA-2011-BLG-274        & 3657.7/3649  & 5742.005      &  0.0029         & 2.65        & 0.0129        & --             \\
                        &              & $\pm$0.001    &  $\pm$0.0001    & $\pm$0.06   & $\pm$0.0003   & --             \\
OGLE-2011-BLG-0990      & 5551.6/5540  & 5758.691      &  0.0151         & 6.70        & 0.0199        & --             \\
/MOA-2011-BLG-300       &              & $\pm$0.001    &  $\pm$0.0004    & $\pm$0.07   & $\pm$0.0003   & --             \\
OGLE-2011-BLG-1101      & 1562.6/1562  & 5823.574      &  0.0485         & 29.06       & 0.0979        & --             \\
/MOA-2011-BLG-325       &              & $\pm$0.002    &  $\pm$0.0005    & $\pm$0.11   & $\pm$0.0006   & --               
\enddata  
\tablecomments{ 
The parameters of the first 6 events are adopted from previously analyses and 
those of the other 9 events are determined in this work.  For OGLE-2004-BLG-254, 
the event was reanalyzed by adding more data sets.  The references of the 
previous analyses are presented in Table \ref{table:one}.  
}
\end{deluxetable*}

\begin{deluxetable*}{lllll}
\tabletypesize{\small}
\tablecaption{Source Parameters\label{table:four}}
\tablewidth{0pt}
\tablehead{
\multicolumn{1}{c}{event} &
\multicolumn{1}{c}{source type} &
\multicolumn{3}{c}{limb-darkening coefficients} \\
\multicolumn{1}{c}{} &
\multicolumn{1}{c}{$(\log g, T_{\rm eff})$} &
\multicolumn{1}{c}{$u_{\it V}$} &
\multicolumn{1}{c}{$u_{\it R}$} &
\multicolumn{1}{c}{$u_{\it I}$}  
}
\startdata
OGLE-2004-BLG-254      & KIII             & --             & 0.70$\pm$0.05  & 0.55$\pm$0.05  \\
OGLE-2004-BLG-482      & MIII             & --             & 0.88$\pm$0.02  & 0.71$\pm$0.01  \\
OGLE-2007-BLG-050      & subgiant         & --             & --             & --             \\
/MOA-2007-BLG-103      &                  &                &                &                \\
OGLE-2007-BLG-224      & FV               & --             & --             & --             \\
/MOA-2007-BLG-163      &                  &                &                &                \\
OGLE-2008-BLG-279      & GV               & --             & --             & --             \\
/MOA-2008BLG-225       &                  &                &                &                \\
OGLE-2008-BLG-290      & KIII             & 0.77$\pm$0.01  & 0.62$\pm$0.07  & 0.55$\pm$0.01  \\
/MOA-2008-BLG-241      &                  &                &                &                \\
\hline                                    
OGLE-2004-BLG-254      & KIII             & --                         & --                              & 0.70$\pm$0.07 (OGLE)           \\  
                       & (2.0, 4750 K)    &                            &                                 & 0.56$\pm$0.10 (CTIO)           \\
                       &                  &                            &                                 & 0.69$\pm$0.10 (Boyden)         \\
                       &                  &                            &                                 & 0.78$\pm$0.09 (Canopus)        \\
                       &                  &                            &                                 & 0.55$\pm$0.06 (SAAO)           \\
                       &                  &                            &                                 & {\bf 0.61 (Claret 2000)}       \\
MOA-2007-BLG-176       & KIII             & --                         & 0.53$\pm$0.04 (MOA)             & 0.50$\pm$0.05 (CTIO)           \\
                       & (2.0, 4500 K)    &                            & 0.51$\pm$0.05 (Auckland)        & 0.44$\pm$0.06 (Lemmon)         \\
                       &                  &                            & {\bf 0.73 (Claret 2000)}        & {\bf 0.63 (Claret 2000)}       \\
MOA-2007-BLG-233       & GIII             & --                         & 0.56$\pm$0.02 (MOA)             & 0.53$\pm$0.04 (OGLE)           \\
/OGLE-2007-BLG-302     & (2.5, 5000 K)    &                            & {\bf 0.68 (Claret 2000)}        & 0.56$\pm$0.02 (CTIO)           \\
                       &                  &                            &                                 & 0.49$\pm$0.02 (Canopus)        \\
                       &                  &                            &                                 & {\bf 0.59 (Claret 2000)}       \\
MOA-2009-BLG-174       & FV               & --                         & --                              & 0.33$\pm$0.02 (CTIO)           \\
                       & (4.5, 6750 K)    &                            &                                 & {\bf 0.46 (Claret 2000)}       \\
MOA-2010-BLG-436       & --               & --                         & 0.52$\pm$0.10 (MOA)             & --                             \\
                       &                  &                            &                                 &                                \\
MOA-2011-BLG-093       & GIII             & 0.69$\pm$0.05 (CTIO)       & 0.55$\pm$0.04 (MOA)             & 0.51$\pm$0.10 (OGLE)           \\
                       & (3.0, 5500 K)    & {\bf 0.70 (Claret 2000)}   & {\bf 0.63 (Claret 2000)}        & 0.58$\pm$0.04 (CTIO)           \\
                       &                  &                            &                                 & 0.51$\pm$0.03 (Canopus)        \\
                       &                  &                            &                                 & {\bf 0.54 (Claret 2000)}       \\
MOA-2011-BLG-274       & GV               & --                         & 0.48$\pm$0.02 (Kumeu)           & --                             \\
                       & (4.0, 6000 K)    &                            & 0.51$\pm$0.01 (Auckland)        &                                \\
                       &                  &                            & {\bf 0.59 (Claret 2000)}        &                                \\
OGLE-2011-BLG-0990     & --               & --                         & --                              & 0.56$\pm$0.04 (OPD)            \\
/MOA-2011-BLG-300      &                  &                            &                                 &                                \\
OGLE-2011-BLG-1101     & KIII             & 0.89$\pm$0.14 (CTIO)       & 0.77$\pm$0.08 (MOA)             & 0.74$\pm$0.07 (OGLE)           \\ 
/MOA-2011-BLG-325      & (2.0, 4250 K)    & {\bf 0.83 (Claret 2000)}   & {\bf 0.76 (Claret 2000)}        & 0.81$\pm$0.07 (CTIO)           \\
                       &                  &                            &                                 & 0.77$\pm$0.06 (Canopus)        \\
                       &                  &                            &                                 & 0.78$\pm$0.05 (FTS)            \\
                       &                  &                            &                                 & {\bf 0.65 (Claret 2000)}                     
\enddata
\tablecomments{ 
The parameters of the first 6 events are adopted from previously analyses and 
those of the other 9 events are analyzed in this work.  For OGLE-2004-BLG-254, 
the event was reanalyzed by adding more data sets.  The references of the 
previous analyses are presented in Table \ref{table:one}. The limb-darkening 
coefficients, $u_{\lambda}$, are presented for the individual data sets used 
for $u_{\lambda}$ measurements and they are compared with theoretical values 
predicted by \citet{claret00}. Also presented are the adopted values of $\log g$ 
and $T_{\rm eff}$. The unit of the stellar surface gravity is cm/s$^2$
}
\end{deluxetable*}

\section{Result}

In Table \ref{table:three}, we present the lensing parameters of the 
best-fit solutions of the individual events determined from modeling. 
To provide integrated results of events with lenses passing over source 
stars, we also provide solutions of events that were previously analyzed.  
For OGLE-2004-BLG-254, we provide both solutions of the previous 
analysis and this work for comparison.

For all events analyzed in this work, we are able to measure the 
limb-darkening coefficients of source stars. In Table \ref{table:four}, 
we present the measured limb-darkening coefficients. We measure the 
coefficients corresponding to the individual data sets covering the peak 
of each light curve instead of the individual passbands. This is because 
the characteristics of filters used for different telescopes are different 
from one another even though they are denoted by a single representative band and 
thus joint fitting of data measured in different filter systems may result in 
erroneous measurement of limb-darkening coefficients \citep{fouque10}. 
To compare with theoretical values, we also provide values of coefficients 
predicted by \citet{claret00} for the Bessell $V$, $R$, and $I$ filters. 
Also provided are the source types and the adopted values of $\log g$ and 
$T_{\rm eff}$ where the typical uncertainties of the surface gravity and 
the effective temperature are $\Delta(\log g)=0.5$ and $\Delta T_{\rm eff}=250$ K, 
respectively. We adopt a solar metallicity. We note that the measured 
coefficients are generally in good agreement with theoretical values, 
$u_{\rm th}$. From the table, it is found that for 23 out of the total 
29 measurements the measured coefficients are within 20\% range of the 
fractional difference as measured by $f_u=(u-u_{\rm th})/u_{\rm th}$. 
The cases of large differences with $f_u>20\%$ include $u_I$(Canopus) 
for OGLE-2004-BLG-254, $u_R$(MOA), $u_R$(Auckland), and $u_I$(Lemmon) 
for MOA-2007-BLG-176, $u_I$(CTIO) for MOA-2009-BLG-174, and $u_I$(CTIO) 
for OGLE-2011-BLG-1101/MOA-2011-BLG-325. From the inspection of the individual 
data points on the light curves, we find the major reasons for the large 
differences between the measured and theoretical values are due to poor 
coverage [$u_I$(Canopus) for OGLE-2004-BLG-254, $u_R$(MOA, Auckland) and 
$u_I$(Lemmon) for MOA-2007-BLG-176, $u_I$(CTIO) for MOA-2009-BLG-174] or 
poor data quality [$u_I$(CTIO) for OGLE-2011-BLG-1101/MOA-2011-BLG-325]. 
Other possible reason for differences from the predicted values include 
differences of individual filters from the standard Bessel filters, 
as well as differences in the method to compute the theoretical values 
\citep{heyrovsky07}.

\begin{figure*}[ht]
\epsscale{0.8}
\plotone{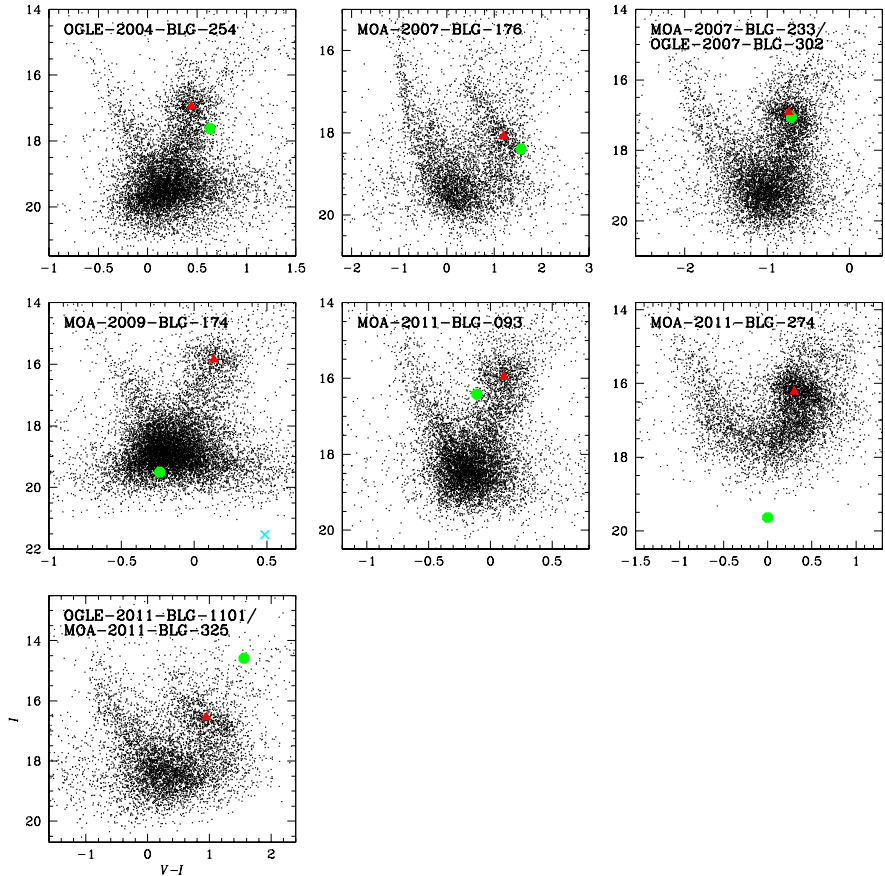}
\caption{\label{fig:ten}
Color-magnitude diagrams of neighboring stars in the fields of lensing events.
In each panel, the circle represents the location of the lensed star 
and the triangle is the centroid of clump giants that is used as a reference
for color and brightness calibration. For MOA-2009-BLG-174, 
the `X' mark denotes the location of the blend.      
}\end{figure*}

The source type of each event is determined based on the location of the 
source in the color-magnitude diagram (CMD) of stars in the same field. 
CMDs are obtained from CTIO images taken in ${\it V}$ and ${\it I}$ 
bands. To locate the lensed star in the CMD, it is required to measure 
the fraction of blended light in the observed light curve. This is done 
by including a blending parameter in the process of light curve modeling. 
For MOA-2011-BLG-274, a CMD taken from CTIO is available but images were 
taken after the event and thus we could not determine the source color and 
magnitude by the usual method. Instead we employ the method of \citet{gould10}. 
In this method, we first measure the source instrumental magnitudes 
by fitting the OGLE ($I_{\rm OGLE}$) and PEST (unfiltered, $N_{\rm PEST}$) 
data to the light curve model. We then align each of these data sets to CTIO ($V/I$) 
using comparison stars, which effectively transforms  $N_{\rm PEST}/I_{\rm OGLE}$ to
$(V/I)_{\rm CTIO}$. In figure \ref{fig:ten}, we present the CMDs of stars 
in the fields of the individual events and the locations of source stars.\footnote{We 
note that high-resolution spectra are available for some events 
with lenses passing over source stars. These events are OGLE-2004-BLG-254 
\citep{cassan06}, OGLE-2004-BLG-482 \citep{zub11}, OGLE-2007-BLG-050/MOA-2007-BLG-103 
\citep{johnson11}, MOA-2009-BLG-174, MOA-2010-BLG-311, MOA-2010-BLG-523 
\citep{bensby11}, and MOA-2011-BLG-093 \citep{mcgregor11}. For those who 
are more interested in the source stars of these events, see the related 
references.}  For MOA-2010-BLG-436 and OGLE-2011-BLG-0990/MOA-2011-BLG-300,
there exists SAAO data taken in {\it I} and {\it V} bands, but the number 
and quality of {\it V}-band data are not numerous and good enough to specify 
the source type.

In Table \ref{table:five}, we present the measured Einstein radii. The 
Einstein radius of each event is determined from the angular source radius, 
$\theta_{\star}$, and the normalized source radius, $\rho_{\star}$, as 
$\theta_{\rm E}=\theta_{\star}/\rho_{\star}$. The normalized source radius 
is measured from modeling. To measure the angular source radius, we use the 
method of \citet{yoo04}, where the de-reddened ${\it V}-{\it I}$ color is measured 
from the location of the source in the CMD, ${\it V}-{\it I}$ is converted into 
${\it V}-{\it K}$ using the relation of \citet{bessel88}, and then the angular 
source radius is inferred from the ${\it V}-{\it K}$ color and the surface 
brightness relation given by \citet{kervella04}. In this process, we use 
the centroid of bulge clump giants as a reference for the calibration of 
the color and brightness of a source under the assumption that the source 
and clump giants experience the same amount of extinction and reddening. 
We note that no CMD is available for MOA-2010-BLG-436 and 
OGLE-2011-BLG-0990/MOA-2011-BLG-300 and thus the Einstein radius 
is not provided. Also provided in Table \ref{table:five} are 
the relative lens-source proper motions as measured by $\mu=\theta_{\rm E}/t_{\rm E}$.

\begin{deluxetable*}{lllll}
\tablecaption{Physical Lens Parameters\label{table:five}}
\tablewidth{0pt}
\tablehead{
\multicolumn{1}{c}{event} &
\multicolumn{1}{c}{$\theta_{\rm E} \left({\rm mas}\right)$} &
\multicolumn{1}{c}{$\mu \left({\rm mas}~{\rm yr}^{-1}\right)$} &
\multicolumn{1}{c}{$M \left(M_{\sun}\right)$} &
\multicolumn{1}{c}{$D_{\rm L} \left({\rm kpc}\right)$} 
}
\startdata
OGLE-2004-BLG-254                     & 0.114            & 3.1            & --               & --             \\
OGLE-2004-BLG-482                     & 0.4              & 16             & --               & --             \\
OGLE-2007-BLG-050/MOA-2007-BLG-103    & 0.48$\pm$0.01    & 2.63$\pm$0.08  & 0.50$\pm$0.14    & 5.5$\pm$0.4    \\
OGLE-2007-BLG-224/MOA-2007-BLG-163    & 0.91$\pm$0.04    & 48$\pm$2       & 0.056$\pm$0.004  & 0.53$\pm$0.04  \\
OGLE-2008-BLG-279/MOA-2008-BLG-225    & 0.81$\pm$0.07    & 2.7$\pm$0.2    & 0.64$\pm$0.10    & 4.0$\pm$0.6    \\
OGLE-2008-BLG-290/MOA-2008-BLG-241    & 0.30$\pm$0.02    & 6.7$\pm$0.4    & --               & --             \\
\hline
OGLE-2004-BLG-254                     & 0.14$\pm$0.01    & 4.06$\pm$0.35  & --               & --             \\
MOA-2007-BLG-176                      & 0.14$\pm$0.01    & 6.21$\pm$0.54  & --               & --             \\
MOA-2007-BLG-233/OGLE-2007-BLG-302    & 0.17$\pm$0.01    & 3.81$\pm$0.33  & --               & --             \\
MOA-2009-BLG-174                      & 0.43$\pm$0.04    & 2.40$\pm$0.24  & 0.84$\pm$0.37    & 6.39$\pm$1.11  \\
MOA-2010-BLG-436                      & --               & --             & --               & --             \\
MOA-2011-BLG-093                      & 0.07$\pm$0.01    & 1.80$\pm$0.16  & --               & --             \\
MOA-2011-BLG-274                      & 0.08$\pm$0.01    & 11.18$\pm$0.97 & --               & --             \\
OGLE-2011-BLG-0990/MOA-2011-BLG-300   & --               & --             & --               & --             \\
OGLE-2011-BLG-1101/MOA-2011-BLG-325   & 0.24$\pm$0.02    & 2.99$\pm$0.26  & --               & --             

\enddata  
\tablecomments{ 
The parameters of the first 6 events are adopted from previously analyses and 
those of the other 9 events are analyzed in this work.  For OGLE-2004-BLG-254, 
the event was reanalyzed by adding more data sets.  The references of the 
previous analyses are presented in Table \ref{table:one}.
}
\end{deluxetable*}

\begin{figure}[ht]
\epsscale{1.1}
\plotone{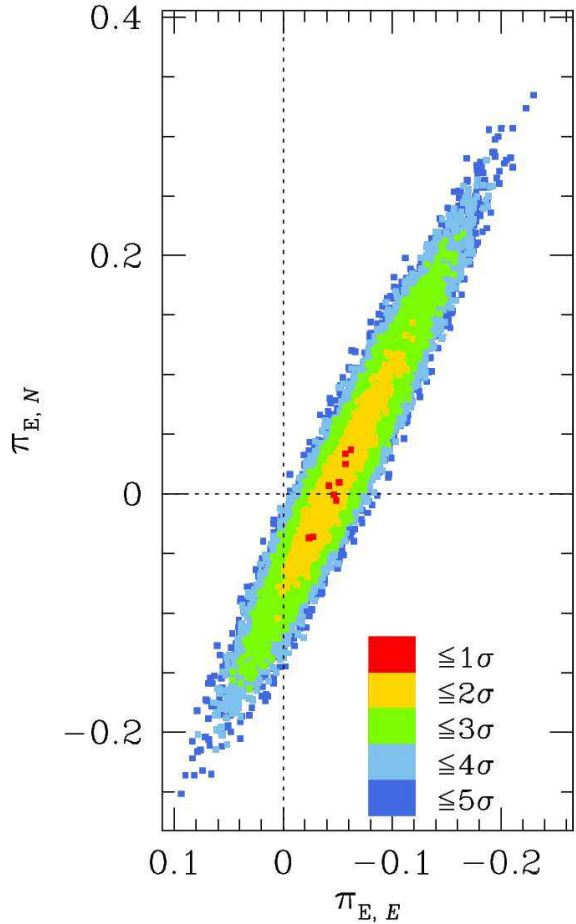}
\caption{\label{fig:eleven}
Contours of $\chi^2$ from the best-fit solution in the space of the parallax 
parameters of the event MOA-2009-BLG-174.
}\end{figure}

\begin{figure*}[ht]
\epsscale{1.1}
\plotone{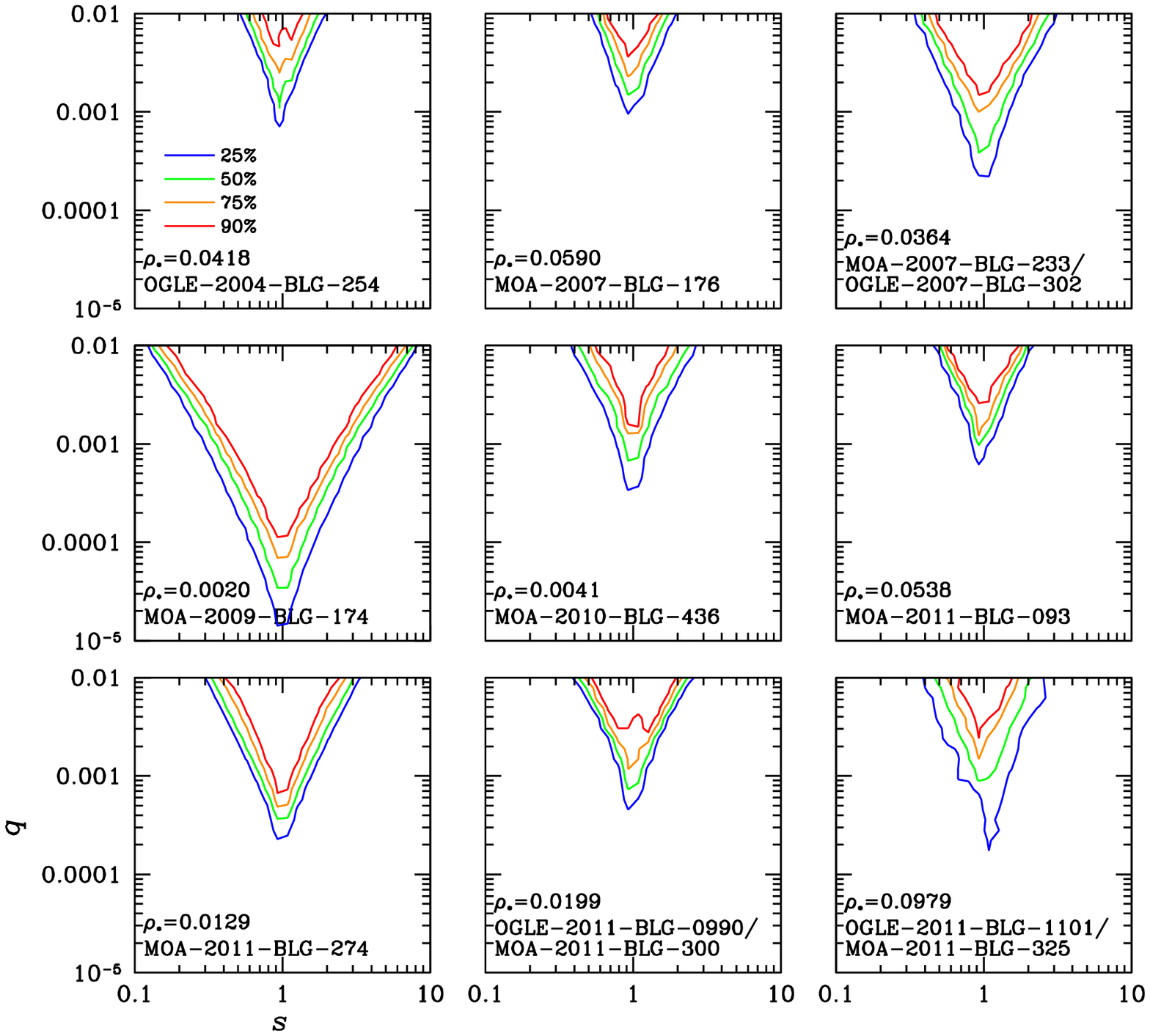}
\caption{\label{fig:twelve}
Exclusion diagrams of planets as a function of the star-planet separation 
(normalized in the Einstein radius) and the planet/star mass ratio.
}\end{figure*}

We note that the measured Einstein radii of some events are substantially 
smaller than a typical value. These events include OGLE-2004-BLG-254 
($\theta_{\rm E}\sim 0.14~{\rm mas}$), MOA-2007-BLG-176 
($\sim 0.14~{\rm mas}$), MOA-2007-BLG-233/OGLE-2007-BLG-302 
($\sim 0.17~{\rm mas}$), MOA-2011-BLG-093 ($\sim 0.07~{\rm mas}$), and 
MOA-2011-BLG-274 ($\sim 0.08~{\rm mas}$).
The lens mass and distance are related to the Einstein radius by
\begin{equation}
M = 0.019~M_\odot
\biggl({D_{\rm S}\over 8~{\rm kpc}}\biggr)
\biggl({D_{\rm L}\over {D_{\rm S} - D_{\rm L}}}\biggr)
\biggl({\theta_{\rm E}\over 0.14~{\rm mas}}\biggr)^2
\label{eq5}
\end{equation}
Hence, the small $\theta_{\rm E}$ of these events implies that lenses 
are either very close to the source or very low-mass objects. Most of 
these events have proper motions that are typical of bulge lenses 
(2--7 $\rm mas\,yr^{-1}$) and so may be quite close to the source 
(see Table 5). But MOA-2011-BLG-274 has a substantially higher proper motion, 
$\mu\sim11\ \rm mas\ yr^{-1}$. It is therefore a good candidate for 
a sub-stellar object or even a free-floating planet \citep{sumi11}. 
Because of its high proper motion, it should be possible to detect the lens 
within a few years using high-resolution infrared imaging, provided 
it is luminous.  In this case a null result would confirm its substellar nature.

For MOA-2009-BLG-174, the lens parallax is measured with $\Delta\chi^2\sim
16.2$.  The measured parallax parameters are
\begin{equation}
\pi_{{\rm E},\parallel}=-0.049\pm0.006~;\qquad \pi_{{\rm E},\perp}=0.038\pm0.065,
\end{equation}
where $\pi_{{\rm E},\parallel}$ and $\pi_{{\rm E},\perp}$ are the components 
of the lens parallax vector that are parallel with and perpendicular to the 
projected position of the Sun. These values correspond to the standard parallax 
components of $(\pi_{{\rm E},N},\pi_{{\rm E},E})=(0.025\pm0.052,-0.057\pm0.028)$.
In Figure \ref{fig:eleven}, we present contours of $\chi^2$ in the space of 
the parallax parameters. Combined with the measured Einstein radius, 
the physical parameters of the lens are uniquely determined as
\begin{equation}
M={{\theta_{\rm E}}\over{\kappa\pi_{\rm E}}}=0.84\pm0.37\ M_{\sun},
\label{eq6}
\end{equation}
and
\begin{equation}
D_{\rm L}={{\rm AU}\over{\pi_{\rm E}\theta_{\rm E}+\pi_{\rm S}}}
=6.39\pm1.11\ {\rm kpc},
\label{eq7}
\end{equation}
respectively. We find that the measured lens mass is consistent with the 
de-reddened color of blended light $\left({\it V}-{\it I}\right)_{0,b}\sim 1.4$, 
which approximately corresponds to the color of an early K-type main-sequence 
star with a mass equivalent to the estimated lens mass, suggesting that the blend 
is very likely to be the lens. We mark the position of the blend in the 
corresponding CMD in Figure \ref{fig:ten}.

A high-magnification event is an important target for planet search due to
its high efficiency to planetary perturbations. Unfortunately, we find no 
statistically significant deviations from the single-lens fit for any of 
the events analyzed in this work. However, it is still possible to place 
limits on the range of the planetary separation and mass ratio. For this 
purpose, we construct so-call ``exclusion diagrams'' which show the confidence 
levels of excluding the existence of a planet as a function of the normalized 
star-planet separation and the planet/star mass ratio. We construct diagrams 
by adopting \citet{gaudi00} method. In this method, binary models are fitted 
to observed data with the 3 binary parameters ($s$, $q$, $\alpha$) are held fixed. 
Then, the confidence level of exclusion for planets with $s$ and $q$ is estimated 
as the fraction of binary models not consistent with the best-fit single-lens 
model among all tested models with various values of $\alpha$. For fitting 
binary models, it is required to produce many light curves with finite magnifications. 
We produce light curves by using the ``map-making method'' \citep{dong06}, 
where a magnification map for a given $s$ and $q$ is constructed and light curves 
with various source trajectories are produced based on the map. In Figure 
\ref{fig:twelve}, we present the obtained exclusion diagrams for all analyzed 
events. Here we adopt a threshold of planet detection as $\Delta \chi^2_{\rm th} 
= \chi^2_{\rm s}-\chi^2_{\rm p} = 200$, where $\chi^2_{\rm p}$ and $\chi^2_{\rm s}$ 
represent the $\chi^2$ values for the best-fit planetary and single-lens 
models, respectively. For most events, the constraints on the excluded parameter 
space is not strong mainly due to the severe finite-source effect. However, 
the constraint is strong for MOA-2009-BLG-174 because of the small 
source size ($\rho_{\star} \sim 0.002$) and dense coverage of the peak.

\section{Summary}

We provide integrated results of analysis for 14 high-magnification lensing 
events with lenses passing over the surface of source stars that have been 
detected since 2004.  Among them, 8 events are newly analyzed in this work.
The newly analyzed events are 
MOA-2007-BLG-176, MOA-2007-BLG-233/OGLE-2007-BLG-302, MOA-2009-BLG-174,
MOA-2010-BLG-436, MOA-2011-BLG-093, MOA-2011-BLG-274, 
OGLE-2011-BLG-0990/MOA-2011-BLG-300, and OGLE-2011-BLG-1101/MOA-2011-BLG-325.
Information about the lenses and lensed stars obtained from the analysis 
is summarized as follows.

\begin{enumerate}
\item
For all newly analyzed events, we measure the linear limb-darkening 
coefficients of the surface brightness profile of the source stars.
\item 
For all events with available CMDs of field stars, we measure the 
Einstein radii and the lens-source proper motions. Among them, 5 events 
(OGLE-2004-BLG-254, MOA-2007-BLG-176, MOA-2007-BLG-233/OGLE-2007-BLG-302, 
MOA-2011-BLG-093, and MOA-2011-BLG-274) are found to have Einstein radii 
less than 0.2 mas, making the lenses of the events candidates of very 
low-mass stars or brown dwarfs.  
\item The measured time scale $t_{\rm E} \sim 2.7$ days combined with 
the small Einstein radius of $\sim 0.08$ mas of the event MOA-2011-BLG-274 
suggests the possibility that the lens is a free-floating planet. 
\item 
For MOA-2009-BLG-174, we additionally measure the lens parallax and thus
uniquely determine the physical parameters of the lens.  The measured 
lens mass of $\sim 0.8\ M_\odot$ is consistent with that of a star blended
with the source, suggesting the possibility that the blend comes from the lens.
\item 
We find no statistically significant planetary signals for any of the 
events analyzed in this work. However, it is still possible to place 
constraint on the range of the planetary separation and mass ratio. 
For this purpose, we provide exclusion diagrams showing the confidence 
levels of excluding the existence of a planet as a function of the 
separation and mass ratio.
\end{enumerate}

\acknowledgments 
Work by CH was supported by Creative Research Initiative Program 
(2009-0081561) of National Research Foundation of Korea.
The MOA experiment was supported by JSPS17340074, JSPS18253002,
JSPS20340052, JSPS22403003, and JSPS23340064.
The OGLE project has received funding from the European Research
Council under the European Community's Seventh Framework Programme
(FP7/2007-2013) / ERC grant agreement no. 246678.
Work by BSG and AG was supported in part by NSF grant AST-1103471.
Work by BSG, AG, RWP, and JCY supported in part by NASA grant NNX08AF40G.
Work by JCY was supported by a National Science Foundation Graduate
Research Fellowship under Grant No.\ 2009068160.
CBH acknowledges the support of the NSF Graduate Research Fellowship 
\#2011082275
%
%
%
TS was supported by the grants JSPS18749004, MEXT19015005, and JSPS20740104.
FF, DR and JS were supported by the Communaut{\'e}
fran\c{c}aise de Belgique - Actions de recherche concert{\'e}es -
Acad{\'e}mie universitaire Wallonie-Europe.

\end{document}